\documentstyle[preprint,aps,version2]{revtex}  

\begin{document}

\font\fortssbx=cmssbx10 scaled \magstep2
\hbox to \hsize{
\hskip.5in \raise.1in\hbox{\fortssbx University of Wisconsin - Madison}
\hfill$\vcenter{\hbox{\bf MAD/PH/807}
            \hbox{\bf UCD-93-38}
            \hbox{\bf RAL-93-101}
            \hbox{December 1993}}$ }

\vspace{.5in}

\begin{title}
SINGLE-LEPTON TOP SIGNALS WITH A $b$-TAG\\
AT THE TEVATRON
\end{title}

\author{V.~Barger$^a$, E.~Mirkes$^a$, J.~Ohnemus$^b$, and R.J.N.~Phillips$^c$}

\begin{instit}
$^a$Physics Department, University of Wisconsin, Madison WI 53706, USA\\
$^b$Physics Department, University of California, Davis, CA 95616, USA\\
$^c$Rutherford Appleton Laboratory, Chilton, Didcot, OX11 0QX, UK
\end{instit}


\begin{abstract} \normalsize
Events with a leptonic $W$-decay plus jets should contain
top-quark signals, but a QCD $W+{\,}$jets background must be separated.
We compare transverse $W$-momentum, jet multiplicity, and $b$-tagging
separation criteria, and find that the main background
after tagging comes from mistagging. We illustrate how to extract the
mass $m_t$ via event reconstructions and how to confirm signal purity
by lepton angular distributions.
\end{abstract}

\thispagestyle{emtpy}
\newpage

   Searches for the $t$-quark are intense
at the Fermilab Tevatron $p \bar p$  collider.  The main
characteristics of  $t\bar t$ pair production and decay are well known.
Current lower limits on $m_t$\cite{cdfd0} are well above $M_W + m_b$,
assuming the Standard Model (SM), so all signals contain
$t \bar t \to W^+ b W^- \bar b$.   As an experimental trigger, at least
one $W$ is usually required to decay leptonically
$W \to \ell \nu$ $(\ell = e,\mu)$,
providing a distinctive isolated high-$p_T^{}$ lepton and
large missing-$p_T^{}$ (denoted $\overlay /p_T^{}$); but large
backgrounds, especially from QCD production of
\hbox{$W+{\,}$jets,} remain to be
separated.  If the second $W$ also decays leptonically and one
$b$-jet is tagged,  the resulting $\ell \ell' b$  signal
may be comparatively clean, but the event rate is
not large and the top mass cannot be directly reconstructed due to
missing neutrinos \cite{baer,han}.

   The advantages of single-lepton top signals, where the second $W$
decays hadronically  $W \to q \bar q'$, are both larger event rate and
direct top mass reconstructibility; the draw-back here is the inherent
uncertainty in QCD $W+{\,}$jets background calculations at the parton level,
especially for high jet multiplicity $n_j$.  Some way must
therefore be found to control or eliminate this background.  Since the
top signals contain four hard partons $b \bar b q \bar q'$ while the QCD
background has typically low $n_j$ and few $b$-quarks,
the usual approach is to require large $n_j$ and $b$-tagging of at least
one jet; calculations then predict that the background is severely reduced
relative to the signal\cite{berends}, assuming this background comes
mainly from $Wb\bar b+{\,}$jets production with genuine $b$-jets.

  In the present Letter we point out
that the main background actually comes from
mistagged events containing no true $b$-jets.
We find that the transverse momentum
$p_T^{}(W)$ of the trigger $W$ is another important characteristic,
and investigate the interplay of $p_T^{}(W),\; n_j$, and $b$-tagging
criteria in separating single-lepton top signals from $W+{\,}$jets
backgrounds.
  The decay  $t \to b W$
has a Jacobian peak at $p_T^{}(W) = (m_t^2 - M_W^2)/(2m_t)$ in the
$t$-restframe, giving a broad $p_T^{}(W)$ lab-frame distribution
unlike the QCD background.   We find that the background
$p_T^{}(W)$-dependence differs less from the signal after imposing
$n_j \ge 3$; nevertheless a $p_T^{}(W)$ cut can be helpful for heavier
top $m_t \agt 170$~GeV.  Once selection cuts have been imposed, we
illustrate how the mass $m_t$ can be found by event reconstructions, with
fitting criteria that further suppress backgrounds, and show how the
signal purity can eventually be confirmed by lepton decay distributions.
Our conclusions are detailed in (i)--(viii) below.

   Analytic next-to-leading order calculations\cite{ark} of inclusive $W$
production agree well with CDF data\cite{cdf1} but cannot address
jet multiplicity with specific acceptance cuts, and anyway exist
only for $n_j \le 2$. We therefore make Monte Carlo parton-level
calculations of $W+n$-jet backgrounds at leading order
\cite{berends,wisc,mangano},  interpreting final partons as jets if
they satisfy the cuts, and imposing typical acceptance cuts:
\begin{equation}
 p_T^{}(\ell,\rm jet, missing) > 20~GeV,\quad
 |\eta(\ell)| < 1.1$,\quad $|\eta(j)| < 2.0,\quad
 \Delta R(jj, j\ell) > 0.4.
\end{equation}
Here $\eta = \ln\tan(\theta/2)$ is pseudo-rapidity,
$(\Delta R)^2 = (\Delta \phi)^2 + (\Delta \eta)^2$, and $\theta$ and
$\phi$ are polar and azimuthal angles relative to the beam.  The
$\Delta R$ cuts approximate some effects of jet-finding and
lepton-isolation criteria.  We assume that at least one of the final
jets is $b$-tagged by a vertex detector (neglecting additional
tagging via semileptonic decays, since the extra neutrino would blur
reconstructions of $t \to Wb$ kinematics).
For $W+{\,}$jets production,  we neglect quark masses (valid at high $p_T^{}$)
and use the scale $Q = \left<p_T^{}(j)\right>$ with the
MRS set D0 parton distributions\cite{mrs} and 4 flavors.  The signals
from $t \bar t$ production and decay are calculated at lowest order,
without $t$-fragmentation effects because of the short top lifetime.
However, we normalize the cross section to $O(\alpha_s^3)$
calculations, taking central values from Ref.\cite{laenen}
(similar central values are given in Ref.\cite{ellis}).   To all
calculations we add gaussian lepton- and jet-smearing
prescriptions\cite{bop}, based on CDF values\cite{smear}, and evaluate
$\overlay / p_T$ from the overall $p_T$ imbalance.

   In the CDF experiment, the efficiency for tagging one or more
$b$-jets in a $t \bar t$ event is about 0.30, corresponding to a
probability $\epsilon_b \simeq 0.16$ per $b$-jet; the probability of a fake
$b$-tag is estimated to be $\epsilon_q \simeq \epsilon_g \simeq 0.01$ per
light-quark or gluon jet\cite{frisch}. We assume a probability
$\epsilon_c \simeq 0.03$ for a bogus $c$-jet tag.  The cross section for
each final configuration is multiplied by the corresponding probability
that at least one of the jets is tagged; e.g. the tagged cross sections
for $Wgggg$, $Wc\bar cgg$, $Wb\bar b q q'$ production contain factors
0.04, 0.08, 0.32, respectively.

   Tagged signal and background cross sections, for separate jet
multiplicities $n_j$, are shown  versus $p_T^{}(W)$ in Fig.~1(a).
Solid curves denote the $n_j = 3$ and $n_j=4$ signals for
the case $m_t = 150$~GeV ($n_j \le 2$ signals are negligible).
Dashed curves show total $n_j=1,2,3,4$ backgrounds from $W+{\,}$jets.
For comparison, the contribution from $Wb\bar bjj$ final states,
containing two true $b$-jets \cite{berends,mangano}, is shown by a
dash-dotted curve.  Figure~1(a) shows that\\
(i) The backgrounds from $Wb\bar b+{\,}$jets channels that contain
    genuine $b$-jets and have attracted most attention
    \cite{berends,mangano}, contribute much less than fake-tags.  If
    cleaner tagging becomes possible, better background suppression
    will follow.\\
(ii) The $W+{\,}$jets background has narrower  $p_T^{}(W)$ dependence than
    the $t\bar t$ signal; the signal gets broader as $m_t$ increases.\\
(iii) Higher-multiplicity background components differ less sharply
    from the signal in their $p_T^{}(W)$ dependence.

   Figure~1(b) compares integrated cross sections above
a minimum cutoff  $p_T^{}(W) > p_T^{\rm min}$, for multiplicities
$n_j = 3,4$.   Dashed curves again denote background, solid
(dotted) curves denote signals for $m_t=150\ (170)$~GeV; in each
case the lower curve refers to $n_j=4$ and the upper curve refers
to the combined $n_j=3,4$ cross section.  These results point to further
conclusions:\\
(iv) For any given $p_T^{\rm min}$ and luminosity, $n_j=4$ is always
    the best choice.  Adding $n_j=3$ to $n_j=4$ data gives more
    signal events $S$ but much more background $B$, such that both
    $S/B$ and the statistical significance $S/\sqrt B$ are decreased.\\
(v) For $m_t \alt 150$~GeV  with $n_j=4$, $p_T^{\rm min}$ cuts confer
    little advantage; they improve $S/B$ but reduce $S/\sqrt B$,
    leaving cleaner but less significant signals.\\
(vi) For heavier top, the broader signal can justify a $p_T^{\rm min}$
    cut; e.g.\ for $m_t=170$~GeV and $n_j=4$, a cut $p_T^{}(W) > 50$~GeV
    increases $S/B$ by $25\%$ with no loss of significance.
    Greater advantages accrue for yet larger $m_t$ .

   With a selected class of 4 jet events with a single-lepton and a $b$-tag,
which are dominantly from $t \bar t$, there are various ways to extract
$m_t$ (see e.g.\ Ref.\cite{baer,berends}); we illustrate three.\\
(a)  We can infer the neutrino longitudinal momentum from the
$W\to \ell \nu$  mass shell constraint, within a two-fold
ambiguity, assuming ${ \bf p}_T^{}(\nu) = \overlay/{ \bf p}_T^{}$.
Then the distribution of invariant mass
$m(\ell \nu b_{\rm tag})$ has a peak at $m_t$ on a
combinatorial background.  Each of the two $W$-solutions is counted
independently; also, for multi-tagged events each tagged jet
contributes independently to this distribution. It
is illustrated in Fig.~2(a) for
$m_t=150$~GeV and $n_j=4$. \\
(b)  Another approach is to identify two final untagged jets arising
from $W\to jj$, satisfying an approximate mass-shell constraint
\begin{equation}
    |m(jj) - M_W| < 15\ \hbox{\rm GeV}.
\end{equation}
Then the tagged jet $b_1$ and the remaining fourth jet $b_2$ are both
presumably $b$-jets (in the desired $t \bar t$ events), and the
distributions of invariant mass $m(jjb_1)$ and $m(jjb_2)$ each
contain a peak at $m_t$  on a combinatorial background (that is smaller
because no two-fold ambiguity is present in $W\to jj$).  This
approach is illustrated in Fig.~2(b),  for $m_t=150$~GeV
and $n_j=4$; the distributions $m(jjb_1)$ and $m(jjb_2)$ are
summed, giving 2 counts per event.  It makes no difference here whether
one or both of the non-$W$ jets are tagged; both are regarded as
$b$-jets.  \\
(c) A better method is to reconstruct both leptonic and hadronic $W$'s.
There are then 4 ways to pair these $W$'s (one $W$ still has the two-fold
ambiguity) with the two remaining jets (presumed to be $b$ and $\bar
b$); each pairing gives two top masses $m_{t1}$ and $m_{t2}$. We
select the pairing in which  $m_{t1}$ and $m_{t2}$ are closest,
subject to a reasonable limit
\begin{equation}
      |m_{t1} - m_{t2}| < 50\ \hbox{\rm GeV},
\end{equation}
and their mean value defines the reconstructed top mass $\tilde m_t$:
\begin{equation}
      \tilde m_t = (m_{t1} + m_{t2})/2.
\end{equation}
The two-fold $W \to \ell \nu$ ambiguity and pairing ambiguities are
thus resolved\cite{bop} and a sharper $t$-mass peak results,
as shown in Fig.~2(c) for $m_t=150$~GeV with $n_j=4$. The integrated
$t \bar t$ signal here is 0.22 pb and the background is 0.017 pb,
corresponding to 4.6 events on a background of 0.4 events with
the present accumulated luminosity 21~pb$^{-1}$ at CDF.
For $m_t=170$~GeV the signal is 0.12 pb.
Figure~2 indicates a further conclusion:\\
(vii) Full reconstruction as in (c) gives the cleanest and narrowest
peak, hence best $m_t$ resolution.\\
We note that a different kind of approach is to use
a maximum-likelihood analysis on individual signal  events\cite{kondo},
with the background suppressed by  tagging with $p_T(W)$ and/or
$n_j$  cuts.

   The final event sample, after all cuts, can be examined to confirm
the characteristics expected of a $t \bar t$ signal. First, the $p_T(W)$
distribution should agree with Fig.~1. Second, the charged-lepton
rapidity distribution should be forward/backward symmetric, unlike
QCD $W$-events; Fig.~3(a) compares the asymmetry
$A(y_{\ell}) = \pm
[d \sigma/dy(y_{\ell}) - d \sigma/dy(-y_{\ell})]/
[d \sigma/dy(y_{\ell}) + d \sigma/dy(-y_{\ell})]$
for charged leptons. (The $\pm$ sign is equal to the lepton's charge
and $y > 0$ is the hemisphere in the $p$ beam direction.)
Also decay distributions in the Collins-Soper frame\cite{cs} (where
the $W \to \ell \nu$ reconstruction gives simply a $\pm \cos \theta$
ambiguity), offer further distinctive differences between signal and
background\cite{mirkes}; see Figs.~3(b) and 3(c). There are shape
differences in both $\phi$ and $|\cos \theta |$ distributions,
especially the latter.   We conclude \\
(viii) Charged lepton distributions offer additional purity checks on
a selected $t \bar t$ signal.

\acknowledgments
VB and RJNP thank Peter Landshoff and John
Taylor for the hospitality of DAMTP, Cambridge, where some of
this work was done. This research was supported in part by the U.S.~Department
of Energy under Contracts No.~DE-AC02-76ER00881 and No.~DE-FG03-91ER40674,
in part by the Texas National
Research Laboratory Commission under Grants No.~RGFY93-221 and No.~RGFY93-330,
and in part by the
University of Wisconsin Research Committee with funds granted by the Wisconsin
Alumni Research Foundation.

\newpage
\section*{Figure Captions}

\begin{enumerate}

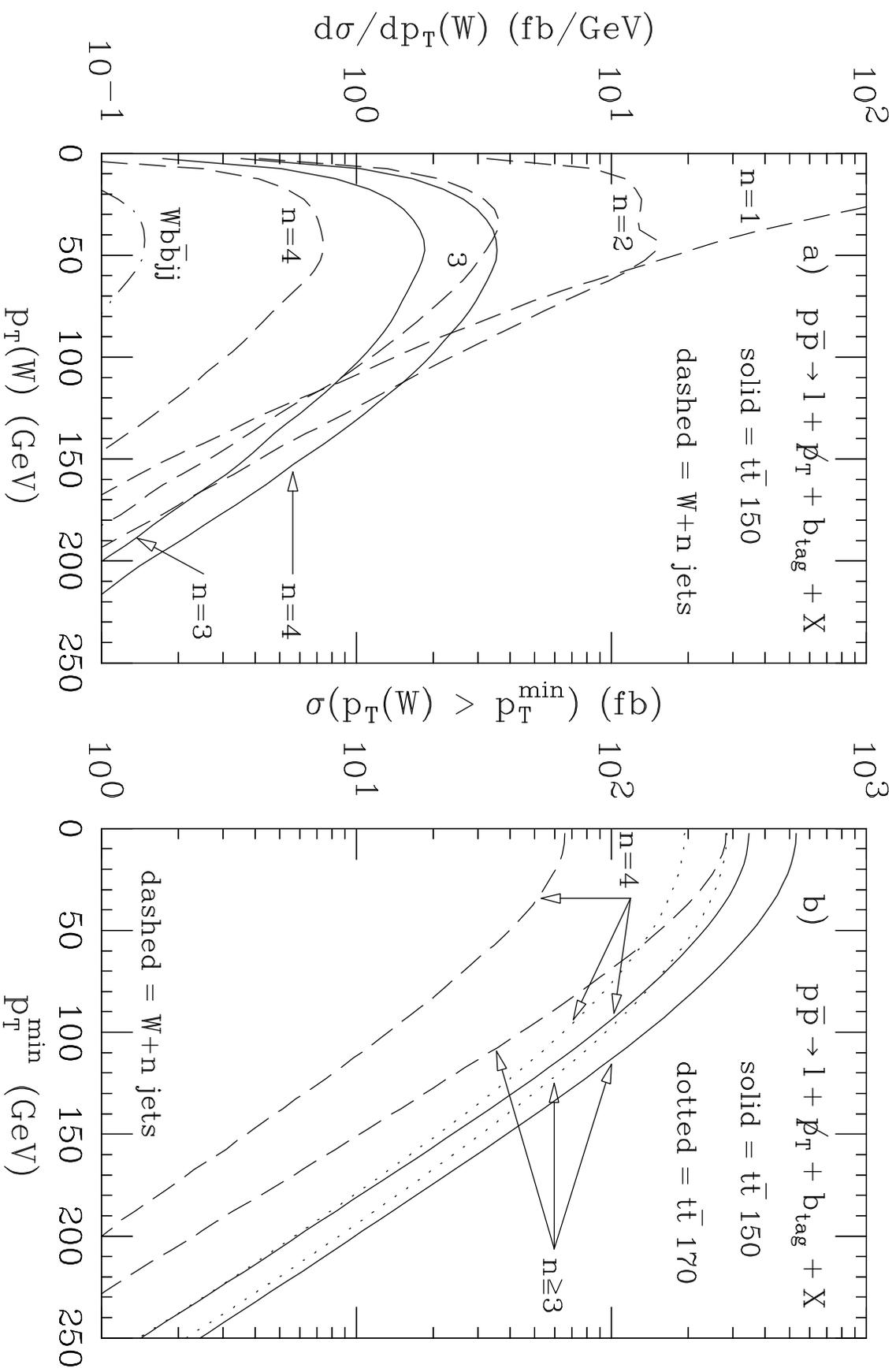
\item[Fig.~1.] $p_T^{}(W)$ and $n_j$ characteristics of $b$-tagged $t\bar t$
signals (solid and dotted curves),
total $W+{\,}$jets backgrounds (dashed curves), and
the background contribution from $Wb\bar bjj$ events (dash-dotted curve):
(a)~differential cross sections versus $p_T(W)$ for various jet
multiplicities $n_j$, (b)~integrated cross sections for
$p_T(W) > p_T^{\rm min}$.

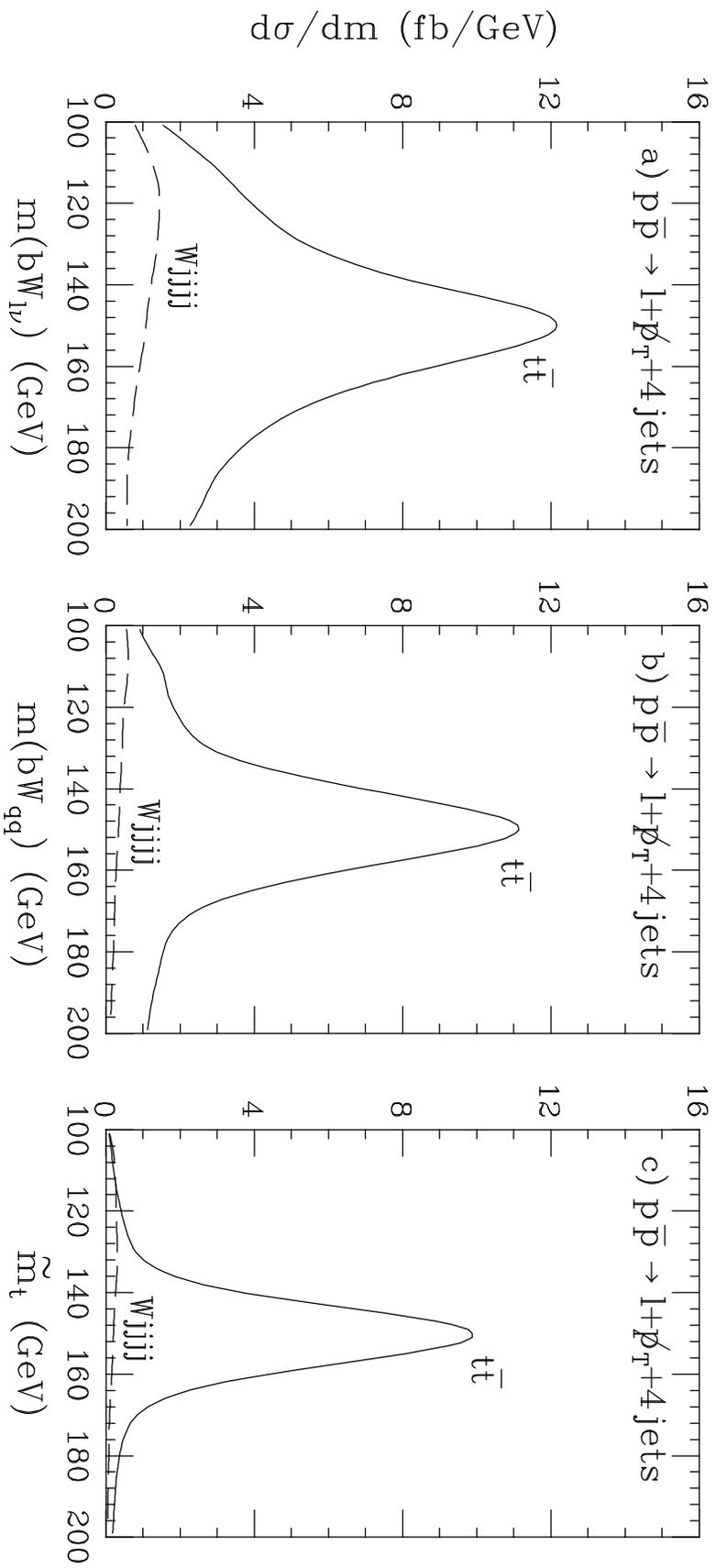
\item[Fig.~2.] Illustrations of three invariant mass reconstructions
described in the text, for the
case $m_t = 150$~GeV with $n_j =4$: (a)~leptonic $W + b$,
(b)~hadronic $W + b$, (c)~best fit to $t \bar t$ kinematics.

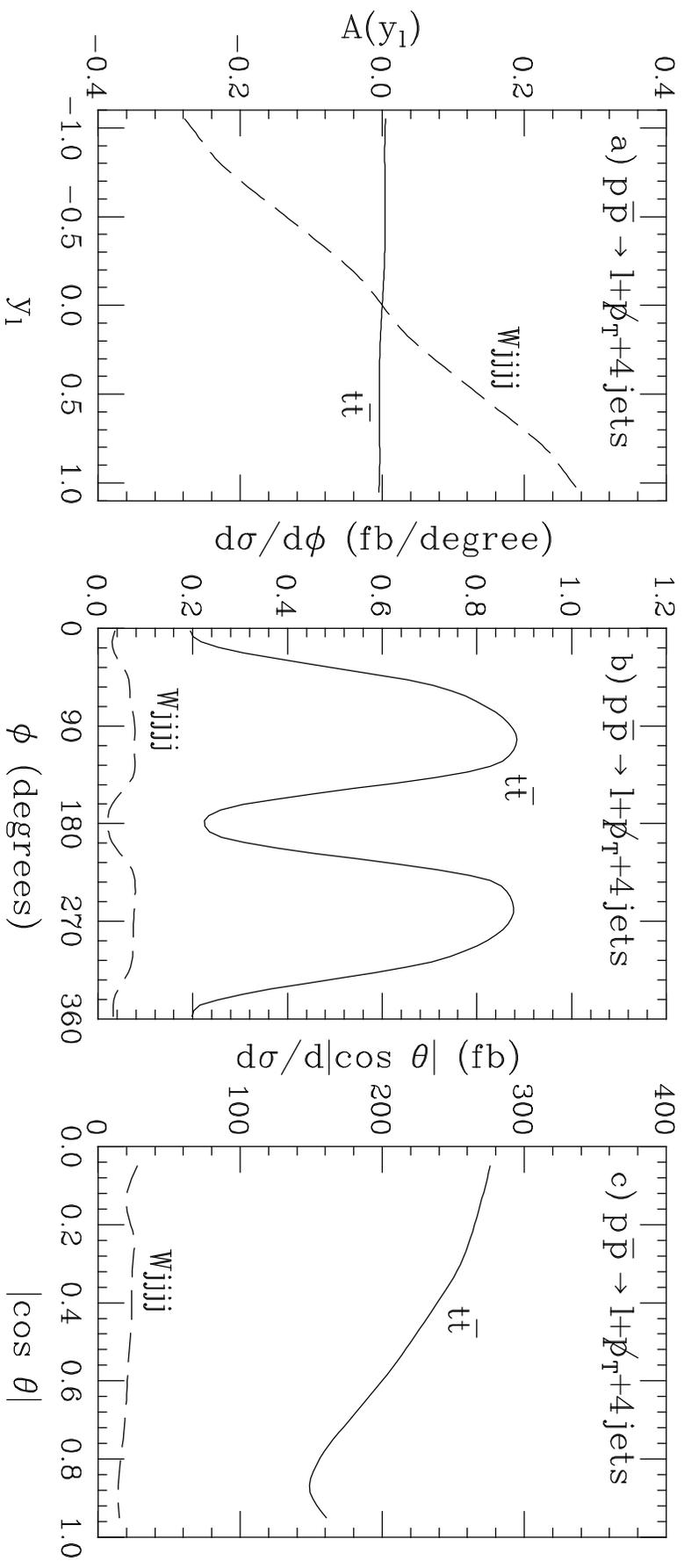
\item[Fig.~3.] Angular distributions of  charged leptons, after
cuts: (a) forward/backward asymmetry versus lepton rapidity $y_{\ell}$,
(b) Collins-Soper azimuthal dependence, (c) Collins-Soper $|\cos \theta |$
dependence.  Solid (dashed) curves denote $t \bar t$ signals
($W+{\,}$jets backgrounds).
\end{enumerate}

\end{document}